\RequirePackage{lineno}
\documentclass[twocolumn,showpacs,preprintnumbers,amsmath,amssymb,superscriptaddress]{revtex4}
\usepackage{graphicx}
\usepackage{dcolumn}
\usepackage{bm}
\modulolinenumbers[1]
\newenvironment{color}[3]
{
}{
}
\newcommand{\grey}[1]     {}

\newcommand{\pT}{$p_{T}$}

\newcommand{\sNN}{$\sqrt{s_{_{NN}}}$}

\newcommand{\DirPho}{$\gamma_{dir}$}
\newcommand{\piZro}{$\pi^{0}$}
\newcommand{\GammaRich}{$\gamma_{rich}$}
\newcommand{\piZroRich}{$\pi^{0}_{rich}$}
\newcommand{\ptAssoc}{$p_{T}^{\mathrm{assoc}}$}
\newcommand{\ptTrig}{$p_{T}^{\mathrm{trig}}$}
\newcommand{\zT}{$z_{T}$}
\newcommand{\DelPhi}{$\Delta\phi$}
\newcommand{\IAApiZro}{$I_{AA}^{\pi^{0}}$}
\newcommand{\IAAg}{$I_{AA}^{\gamma_{dir}}$}
\newcommand{\GeVc}{GeV/$c$}

\begin{document}

\title{ 
Jet-like Correlations with Direct-Photon and Neutral-Pion Triggers at $\sqrt{s_{_{NN}}} = 200$~GeV
}
\affiliation{AGH University of Science and Technology, FPACS, Cracow 30-059, Poland}
\affiliation{Argonne National Laboratory, Argonne, Illinois 60439}
\affiliation{Brookhaven National Laboratory, Upton, New York 11973}
\affiliation{University of California, Berkeley, California 94720}
\affiliation{University of California, Davis, California 95616}
\affiliation{University of California, Los Angeles, California 90095}
\affiliation{Central China Normal University, Wuhan, Hubei 430079}
\affiliation{University of Illinois at Chicago, Chicago, Illinois 60607}
\affiliation{Creighton University, Omaha, Nebraska 68178}
\affiliation{Czech Technical University in Prague, FNSPE, Prague, 115 19, Czech Republic}
\affiliation{Nuclear Physics Institute AS CR, 250 68 Prague, Czech Republic}
\affiliation{Frankfurt Institute for Advanced Studies FIAS, Frankfurt 60438, Germany}
\affiliation{Institute of Physics, Bhubaneswar 751005, India}
\affiliation{Indian Institute of Technology, Mumbai 400076, India}
\affiliation{Indiana University, Bloomington, Indiana 47408}
\affiliation{Alikhanov Institute for Theoretical and Experimental Physics, Moscow 117218, Russia}
\affiliation{University of Jammu, Jammu 180001, India}
\affiliation{Joint Institute for Nuclear Research, Dubna, 141 980, Russia}
\affiliation{Kent State University, Kent, Ohio 44242}
\affiliation{University of Kentucky, Lexington, Kentucky, 40506-0055}
\affiliation{Korea Institute of Science and Technology Information, Daejeon 305-701, Korea}
\affiliation{Lamar University, Beaumont, Texas 77710}
\affiliation{Institute of Modern Physics, Chinese Academy of Sciences, Lanzhou, Gansu 730000}
\affiliation{Lawrence Berkeley National Laboratory, Berkeley, California 94720}
\affiliation{Lehigh University, Bethlehem, Pennsylvania 18015}
\affiliation{Max-Planck-Institut fur Physik, Munich 80805, Germany}
\affiliation{Michigan State University, East Lansing, Michigan 48824}
\affiliation{National Research Nuclear Univeristy MEPhI, Moscow 115409, Russia}
\affiliation{National Institute of Science Education and Research, Bhubaneswar 751005, India}
\affiliation{National Cheng Kung University, Tainan 70101}
\affiliation{Ohio State University, Columbus, Ohio 43210}
\affiliation{Institute of Nuclear Physics PAN, Cracow 31-342, Poland}
\affiliation{Panjab University, Chandigarh 160014, India}
\affiliation{Pennsylvania State University, University Park, Pennsylvania 16802}
\affiliation{Institute of High Energy Physics, Protvino 142281, Russia}
\affiliation{Purdue University, West Lafayette, Indiana 47907}
\affiliation{Pusan National University, Pusan 46241, Korea}
\affiliation{Rice University, Houston, Texas 77251}
\affiliation{University of Science and Technology of China, Hefei, Anhui 230026}
\affiliation{Shandong University, Jinan, Shandong 250100}
\affiliation{Shanghai Institute of Applied Physics, Chinese Academy of Sciences, Shanghai 201800}
\affiliation{State University Of New York, Stony Brook, NY 11794}
\affiliation{Temple University, Philadelphia, Pennsylvania 19122}
\affiliation{Texas A\&M University, College Station, Texas 77843}
\affiliation{University of Texas, Austin, Texas 78712}
\affiliation{University of Houston, Houston, Texas 77204}
\affiliation{Tsinghua University, Beijing 100084}
\affiliation{United States Naval Academy, Annapolis, Maryland, 21402}
\affiliation{Valparaiso University, Valparaiso, Indiana 46383}
\affiliation{Variable Energy Cyclotron Centre, Kolkata 700064, India}
\affiliation{Warsaw University of Technology, Warsaw 00-661, Poland}
\affiliation{Wayne State University, Detroit, Michigan 48201}
\affiliation{World Laboratory for Cosmology and Particle Physics (WLCAPP), Cairo 11571, Egypt}
\affiliation{Yale University, New Haven, Connecticut 06520}

\author{L.~Adamczyk}\affiliation{AGH University of Science and Technology, FPACS, Cracow 30-059, Poland}
\author{J.~K.~Adkins}\affiliation{University of Kentucky, Lexington, Kentucky, 40506-0055}
\author{G.~Agakishiev}\affiliation{Joint Institute for Nuclear Research, Dubna, 141 980, Russia}
\author{M.~M.~Aggarwal}\affiliation{Panjab University, Chandigarh 160014, India}
\author{Z.~Ahammed}\affiliation{Variable Energy Cyclotron Centre, Kolkata 700064, India}
\author{I.~Alekseev}\affiliation{Alikhanov Institute for Theoretical and Experimental Physics, Moscow 117218, Russia}
\author{D.~M.~Anderson}\affiliation{Texas A\&M University, College Station, Texas 77843}
\author{A.~Aparin}\affiliation{Joint Institute for Nuclear Research, Dubna, 141 980, Russia}
\author{D.~Arkhipkin}\affiliation{Brookhaven National Laboratory, Upton, New York 11973}
\author{E.~C.~Aschenauer}\affiliation{Brookhaven National Laboratory, Upton, New York 11973}
% ...

\bigskip

\collaboration{STAR Collaboration}\noaffiliation
 
 \bigskip

\date{ \today / Revised version: v7}
 
\begin{abstract}

Azimuthal correlations of charged hadrons with direct-photon (\DirPho) and 
neutral-pion (\piZro) trigger particles are analyzed
in central Au+Au and minimum-bias $p+p$ collisions at
\sNN~= 200 GeV in the STAR experiment. The charged-hadron per-trigger yields at
mid-rapidity
from central Au+Au collisions are compared with $p+p$ collisions to quantify the suppression in Au+Au collisions. 
The suppression of the away-side associated-particle yields per
\DirPho~trigger is independent of the transverse momentum of the trigger particle (\ptTrig), whereas the suppression is smaller
at low transverse momentum of the associated charged hadrons (\ptAssoc). 
Within uncertainty, similar levels of suppression are observed for \DirPho~and \piZro~triggers as a function of \zT~($\equiv p_T^{\mathrm{assoc}}/p_T^{\mathrm{trig}}$).
The results are compared with energy-loss-inspired theoretical model predictions. Our studies support previous conclusions that the lost energy reappears predominantly at low transverse momentum, regardless of the trigger energy.

\end{abstract}

\maketitle

\section{Introduction}
Over the past decade, experiments at the Relativistic Heavy Ion Collider (RHIC) at BNL
have studied the hot and dense medium created in
heavy-ion collisions. The
suppression of high-transverse momentum ($p_{T}$) inclusive 
hadrons~\cite{suppression_PHENIX,suppression_STAR,suppression_STAR2}, 
indicative of
jet quenching, corroborates the conclusion that the medium created is
opaque to colored energetic partons~\cite{STAR_WP,PHENIX_WP,PHOBOS_WP,BRAHMS_WP}. This phenomenon can be understood
as a result of the medium-induced
radiative energy loss of a hard-scattered parton as it
traverses the Quark Gluon Plasma (QGP) created in heavy-ion
collisions~\cite{Gyulassy_Levai_Vitev,Gyulassy_Levai_Vitev1}. 
The angular correlation of charged hadrons with respect to a direct-photon (\DirPho) trigger 
was proposed as a promising probe to study the mechanisms of parton energy loss~\cite{Wang_Huang_Sarcevic}.  
The presence of a ``trigger'' particle, having \pT~greater than some selected value, serves as part of the selection criteria 
to analyze the event for a hard scattering.
Direct photons are produced 
during the early stage of the collision,
through leading-order pQCD processes such as quark-gluon Compton scattering
($qg\rightarrow q\gamma$) and quark-antiquark pair annihilation ($q\bar{q}\rightarrow g\gamma$). In these processes 
the transverse energy of
the trigger photon approximates the initial \pT~of the outgoing recoil parton,
before the recoiling (``away-side'') parton likely loses energy
while traversing the medium and fragments into a jet. 
The jet-like yields associated with
a trigger particle are estimated by integrating the correlated yields of
charged hadrons over azimuthal distance from the trigger particle ($\Delta \phi$).
Any suppression of the charged-hadron per-trigger yields in the away-side jets in central Au+Au collisions is then quantified by contrasting
to the per-trigger yields measured in $p+p$
collisions, via the ratio of integrated
yields, $I_{AA}$~\cite{PHENIX_GJet,STAR_GJet} (defined in Eq.~\ref{eq:IAA}). 
When requiring a hadron trigger (such as a \piZro), the \pT~of the recoiling parton (and hence the away-side jet) is 
not as well approximated by the transverse energy of the trigger. 
For example, the PYTHIA Monte Carlo simulator~\cite{PYTHIA} shows that, in $p+p$ collisions at \sNN $=$ 200~GeV, a \piZro~trigger with \pT~$>$ 12 \GeVc~carries, on average, only $80\pm5\%$ of the original scattered parton's \pT. This percentage from PYTHIA is consistent with the values extracted from this analysis, as described below.

Despite this complication, it is compelling to compare the suppression for \DirPho~triggers with that for 
\piZro~triggers because of the expected differences in geometrical biases at RHIC energies~\cite{Renk}.
While the \piZro~trigger is likely to have been produced near the surface of the medium, 
the \DirPho~trigger does not suffer the same bias, since the photon mean free path is much
larger than the size of the medium.
Comparing \DirPho- and \piZro-triggered yields
offers further opportunities to explore the 
geometric biases and their interplay with parton energy loss. 
A next-to-leading order perturbative QCD 
calculation~\cite{Zhang_Owens_EWang_XWang_PRL} 
suggests that production of hadrons at different $z_T$ is also affected 
by different geometric biases, where \zT $\equiv p_T^{\mathrm{assoc}}/p_T^{\mathrm{trig}}$ (\ptAssoc~and \ptTrig~are transverse momenta of associated and triggered particles, respectively) 
represents the ratio of the transverse momentum carried by a charged hadron 
in the recoil jet to that of the trigger particle.
The high-\zT~hadrons in a jet recoiling from a $\gamma_{dir}$ preferentially
originate from a parton scattering near the away-side surface of the medium,
since scatterings deeper in the medium will result in a stronger 
degradation of the high-momentum components of the jet.  The high-$z_{T}$ hadrons 
in a jet recoiling from a $\pi^0$ preferentially
emerge from scatterings tangential to the surface from the already biased 
surface-dominated trigger jets (which is consistent with observations in~\cite{STAR_2plus1}).
These two mechanisms turn out to lead to the same level of suppression~\cite{Zhang_Owens_EWang_XWang_PRL}.  
Only at low $z_{T}$ does the
full sampling of the volume by $\gamma_{dir}$ triggers show a 
predicted difference from that 
of the surface-dominated $\pi^{0}$ triggers.

An additional effect at low $z_T$ may be the redistribution of the parton's lost energy within the low-momentum jet fragments~\cite{YAJEM}, which is not included in the
calculation~\cite{Zhang_Owens_EWang_XWang_PRL} described above.  This was studied 
by the PHENIX Collaboration, which found an
enhancement at low $z_T$ and large angles, for direct photon triggers with $p_T^{trig}$ in the range of $5-9$~\GeVc~at \sNN~= 200~GeV in the most central Au+Au collisions~\cite{PHENIX_GJet}.
Enhancements due to this mechanism would be expected in hadrons recoiling from other 
triggers as well, such as $\pi^0$ or jets.  A previous STAR measurement of hadrons associated with a 
reconstructed jet have shown an enhancement for \pT~$ < 2$~\GeVc,
for two classes of jets with broadly separated energy scales, in which the enhancement 
at low $p_T$ balances the suppression at high $p_T$~\cite{STAR_jethadron}.

Furthermore, leading order di-jet production comes from both 
quark and gluon jets.  Recent calculations show that pions with high \pT~relative to the total jet \pT~are predominantly from
quark jets~\cite{deFlorian,Vogelsang}, so, for the jet energies probed in this paper, 
the away-side mainly comes from gluon jets~\cite{DSS07}.
This is in contrast to the
away-side of a \DirPho~trigger, which mainly comes from quark jets, since 
at leading order a photon does not couple with a gluon.
Thus it is expected that, on average, the away-side parton associated with
a \piZro~suffers more energy loss than that of a \DirPho~due to the additional color
factor from gluons. 
By comparing the suppression of
away-side associated hadrons for \DirPho~triggers to that 
for \piZro~triggers, one can gain
information about both the path-length and the color-factor dependence of parton energy loss.

This manuscript is organized as follows.  The detector setup of the STAR experiment is discussed in Sec.~\ref{exp_setup}. 
The transverse shower-shape analysis used to discriminate 
between \piZro~and \DirPho, and the procedures to extract the
charged-hadron spectra, associated with \piZro~and \DirPho~triggers, are discussed in Sec.~\ref{analysis}.
The per-trigger charged-hadron yields are presented as a function of \zT, in Sec.~\ref{results}.  
The dependences of the suppression of these yields in central Au+Au collisions relative to those in minimum-bias $p+p$ collisions on both the trigger energy and the associated transverse momentum are discussed, with comparisons to theoretical model predictions. Finally, 
in Sec.~\ref{summary}, our observations are summarized.

 
\section{Experimental setup}
\label{exp_setup}
The data were taken by the Solenoidal Tracker at RHIC (STAR) experiment
in 2011 and 2009 for Au+Au and $p+p$ collisions at \sNN~= 200 GeV, respectively. 
Using the Barrel Electromagnetic
Calorimeter (BEMC)~\cite{BEMC} to select events 
containing a high-$p_{T}$ $\gamma$ or \piZro, the
STAR experiment collected an integrated 
luminosity of 2.8~nb$^{-1}$ of Au+Au 
collisions and 23~pb$^{-1}$ of $p$+$p$ collisions. 
STAR provides 2$\pi$ azimuthal coverage and wide
pseudo-rapidity ($\eta$) coverage. The Time Projection Chamber
(TPC) is the main charged-particle tracking detector~\cite{TPC}, providing track information for 
the charged hadrons with $|\eta| < 1.0$. The centrality selection is determined from
the charged-particle multiplicity in the TPC within $|\eta| < 0.5$.
 The BEMC is a sampling
calorimeter, and each calorimeter module consists of a lead-scintillator structure and an embedded wire chamber, the Barrel Shower Maximum
Detector (BSMD).  The BSMD is situated approximately five radiation lengths from the front face of the
BEMC. BEMC towers (each covering 0.05 units in $\eta$ and $\phi$) provide a measurement of the energy of
electromagnetic clusters, whereas the BSMD, due to its high
granularity (0.007 units in $\eta$ and $\phi$), provides high spatial
resolution for the center of a cluster and the transverse development of the shower.  
Electromagnetic clusters are constructed from the response of one or two towers, depending on the location of the centroid as
determined by the BSMD.
The transverse extent of the shower is used to distinguish between \DirPho~showers and decay photons from \piZro.  Details of the \piZro/$\gamma$ discrimination are discussed in the next section.

\section{Analysis details}
\label{analysis}
 Events having a transverse energy in a BEMC cluster $E_{T} >$ 8~GeV, with $|\eta|\leq 0.9$, 
are selected for this analysis.
In order to distinguish a \piZro, which at high \pT~predominately decays to
two photons with a small opening angle, from a single-photon cluster, a transverse shower-shape
analysis is performed. In this method, the overall BEMC cluster energy ($E_{cluster}$),
the individual BSMD strip energies ($e_{i}$), and the distances of the
strips ($r_{i}$) from the
center of the cluster are used to construct the ``Transverse Shower Profile'' (TSP).  The TSP is defined as, $\mathrm{TSP} =
{E_{cluster}}/{\sum_{i}e_{i}r_{i}^{1.5}}$~\cite{STAR_GJet,QM15_pro}. 
 The \piZroRich~(nearly pure sample of \piZro) and \GammaRich~(enhanced fraction of
\DirPho) samples
are selected by requiring $\mathrm{TSP}< 0.08$ and $0.2< \mathrm{TSP}<0.6$, respectively,
in both $p+p$ and Au+Au collisions.  The \piZroRich~sample is estimated to be $\sim 95$\% pure \piZro, determined from studies of simulated \piZro~and 
\DirPho~embedded into real data.
The \DelPhi~azimuthal correlations are constructed with charged-hadron tracks within
$1.2$~\GeVc~$<$ \ptAssoc $<$ \ptTrig~and $|\eta|<1.0$.
Both trigger samples are selected with $12<$ \ptTrig $<20$~\GeVc~(or $8<$ \ptTrig $<20$~\GeVc~for the study of the 
\DirPho~\ptTrig~dependence)~and
$|\eta|<0.9$. There is an additional requirement 
that no track with momentum greater than 3~\GeVc~is pointing to the trigger tower. 
This track-rejection cut prevents significant contamination of the measured BEMC 
energy of the trigger particle. The $p_{T}$ threshold of the track-rejection cut was varied between 1 and 4~\GeVc, as a part of 
the systematic studies, and the variations showed no significant difference in the away-side charged-hadron yields.

\begin{figure*}[htbp]
\begin{center}
 \includegraphics[width=0.6\textwidth]{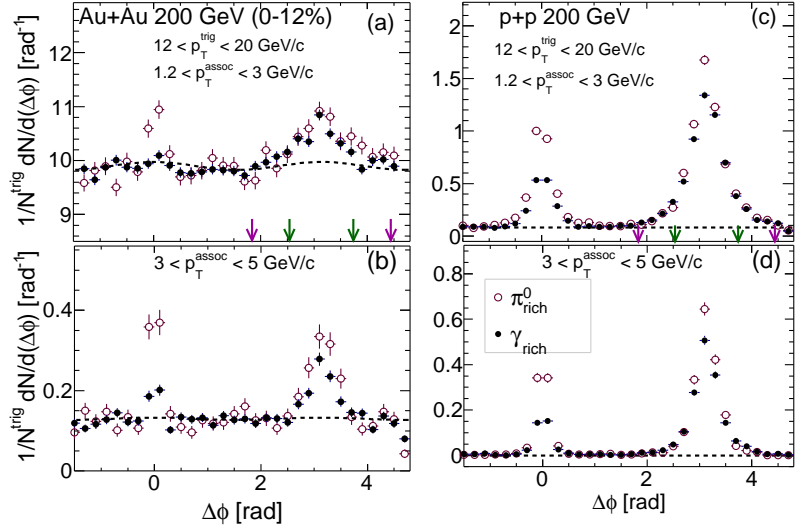}  	
\caption{(Color online.) The azimuthal correlation functions of charged
hadrons per trigger for  $\pi^{0}_{rich}$~(open circles) and  $\gamma_{rich}$~(filled
circles) triggers, measured in central (0-12$\%$) Au+Au collisions and
minimum-bias $p+p$ collisions. Panels (a) and (c) are for Au+Au and $p+p$ collisions, respectively, in
$1.2 <$ \ptAssoc~$< 3$  \GeVc~and panels (b) and (d) are that for $3
<$  \ptAssoc~$< 5$  \GeVc. The dashed curves indicate the background (charged hadrons not correlated with the jet), shown only for $\pi^0_{rich}$ triggers, and the
arrows indicate the range over which the away-side is integrated (green and violet colored arrows represent  $|\Delta\phi-\pi| \leq 0.6$ and $|\Delta\phi-\pi| \leq 1.4$ respectively).}
\label{Fig1}
\end{center}
\end{figure*}

The correlation functions represent the number of associated charged hadrons ($N_{assoc}$) per trigger
particle, $(1/N_{trig}) (dN_{assoc}/d\Delta\phi)$, as a function of
\DelPhi, where $N_{trig}$ is the number of trigger particles.  The yield is integrated over $\Delta\eta = 2$, with no correction
applied for the particle-pair acceptance in $\Delta\eta$.  
In Fig.~\ref{Fig1}, a sample of the azimuthal correlation functions for \GammaRich- and
\piZroRich-triggered associated charged hadrons, for different 
\ptAssoc~ranges, are shown for the $12\%$ most central Au+Au and minimum-bias $p+p$ collisions. In the
lower \ptAssoc~bins, the uncorrelated background (shown in Fig.~\ref{Fig1} as dashed curves) 
is higher than that in higher \ptAssoc~bins, especially in Au+Au collisions, whereas in $p+p$ collisions,
this uncorrelated background is small in all
\ptAssoc~bins.  
On the near-side ($\Delta\phi \sim 0$) the \piZroRich-triggered correlated yields are
larger than those for \GammaRich~triggers, as expected. The non-zero near-side \GammaRich-triggered yields 
are due to the background in the \GammaRich~trigger sample and are used to determine the amount of background, as 
further discussed below.
In the higher \ptAssoc~range, it is also observed that 
the away-side ($\Delta\phi \sim \pi$)
 \GammaRich-triggered yields are smaller than those of the 
\piZroRich~triggers, which can be understood since the 
\piZro~triggers originate from the fragmentation of partons generally having a higher energy than the 
corresponding direct-photon triggers. 

The background subtraction and the pair-acceptance correction (in $\Delta\phi$) have been performed
using a mixed-event technique (see {\it e.g.}~\cite{STAR_2plus1}) for each \zT~bin. 
Event mixing is performed among events having similar vertex position and centrality class.
In Au+Au collisions, the background ({\it i.e.} what is not correlated with the jet) 
may still contain azimuthal correlations due to flow.  The distributions of background pairs for different
\zT~bins are therefore modulated with the second Fourier (elliptic flow) coefficient ($v_{2}$) of the particle
azimuthal distribution measured with respect to the event plane. It
is given by $\mathrm{B[1+2\langle v_{2}^{trig}\rangle\langle v_{2}^{assoc}\rangle cos(2\Delta\phi)]}$,
where $\mathrm{B}$ represents the level of background pairs and is determined assuming applicability of the ``Zero-Yield at 1 radian'' (ZYA1) method, 
a variation
on the ``Zero-Yield at Minimum'' (ZYAM) method~\cite{ZYAM_Ajit}. 
The $\mathrm{\langle v_{2}^{trig}\rangle}$ ($\mathrm{\langle v_{2}^{assoc}\rangle}$) is the average value of the second-order 
flow coefficient~\cite{STAR_v2} of the trigger (associated) particle at the mean \ptTrig~(\ptAssoc) in each $z_T$ bin.
The flow term in the background subtraction only has a significant effect
for Au+Au collisions at low $z_T$, and the higher order flow components are ignored as their magnitudes 
are small in the most central Au+Au collisions. In $p+p$ collisions, $B$ is determined assuming a flat (uncorrelated) background.

The trigger-associated charged-hadron yields are determined from the
azimuthal correlation functions, per trigger particle (\piZroRich~and
\GammaRich~samples), per \DelPhi, both on the near side (\DelPhi~$\sim 0$) and
the away side (\DelPhi~$\sim \pi$). In this analysis, the near-side and
away-side yields are extracted by integrating the correlation
functions, for given \zT~bins, over $|\Delta\phi| \leq 1.4$
and  $|\Delta\phi-\pi| \leq 1.4$, respectively. 
The raw near-side and away-side associated charged-hadron yields
are corrected for the associated-particle efficiencies determined by embedding simulated charged hadrons into real events.
The average tracking efficiencies for charged hadrons (with \ptAssoc~$> 1.2$~\GeVc)
are determined via detector simulations to be around 70$\%$ and 90$\%$ for central Au+Au
and minimum-bias $p+p$ collisions, respectively. 
The \piZro-triggered yields are calculated
from the \piZroRich-triggered correlation functions, with no further correction for the contamination in the trigger sample, 
because of the high purity in the \piZroRich~sample. 
 
Away-side charged-hadron yields for \DirPho~triggers are determined by
assuming zero near-side yield for \DirPho~triggers, and using the following expression
\begin{eqnarray}
 Y_{\gamma_{dir}+h} = \frac{Y^{away}_{\gamma_{rich}+h}-RY^{away}_{\pi^{0}_{rich}+h}}{1-R}.
\label{eq:Yield_dirPhoton}
\end{eqnarray}
Here $Y^{away}_{\gamma_{rich}+h}$ ($Y^{away}_{\pi^{0}_{rich}+h}$) 
represents the away-side yield of \GammaRich~(\piZroRich), and $R$ is given by
\begin{eqnarray}
R = \frac{Y^{near}_{\gamma_{rich}+h}}{Y^{near}_{\pi^{0}_{rich}+h}},
\end{eqnarray}
the ratio of the near-side yield in the $\gamma_{rich}$-triggered correlation function
to the near-side yield in the $\pi^{0}_{rich}$-triggered correlation function.
This means
\begin{eqnarray}
1-R= \frac{N^{\gamma_{dir}}}{N^{\gamma_{rich}}},
\end{eqnarray}
where $N^{\gamma_{dir}}$ ($N^{\gamma_{rich}}$) is the
number of \DirPho~(\GammaRich) triggers. The values of $1-R$, representing 
the fractions of signal in the $\gamma_{rich}$ trigger sample, are found
to be 40$\%$ and 70$\%$ for $p+p$ and the central Au+Au
collisions, respectively.  Using this technique, almost all sources of background 
(including photons from asymmetric hadron decays and fragmentation photons) can be removed, 
assuming that their correlations are similar to those for \piZro~triggers.
This assumption was tested using PYTHIA simulations, 
with decay photons as the trigger particles,
and it was found to be valid to within
at least 15\% (the statistical precision of the PYTHIA study).

Systematic uncertainties include the effects of 
track-quality selection criteria, neutral-cluster selection criteria, 
\piZro/$\gamma$ discrimination (TSP) cuts for the
\piZroRich~and \GammaRich~samples, the size of the 
ZYA1 normalization region, the $v_2$ uncertainty range, and the
yield-integration windows.  All of these sources of uncertainty 
are evaluated for each data point individually. 
For groups of sources that are not independent, such as different yield-extraction conditions, the maximum deviation among the different conditions is taken as the contribution to the systematic error.
The systematic uncertainties from sources that are considered to be independent are added in quadrature.  
The \piZro/$\gamma$ discrimination uncertainty dominates in most \zT~bins, varying between 10 and 25\%.
The track-quality selection criteria typically contributes a 5-10\% uncertainty. 
In the lowest \zT~bin in Au+Au collisions for \piZro~triggers, the yield extraction uncertainty dominates with as much as 50\% uncertainty in the near-side 
yield.
The variation of the $p_T$ threshold for the track-rejection cut for the 
neutral-tower trigger selection typically has a negligible effect.
 
\section{Results and Discussion}
\label{results}
\begin{figure}[htbp]
\begin{center}
  \includegraphics[width=0.5\textwidth]{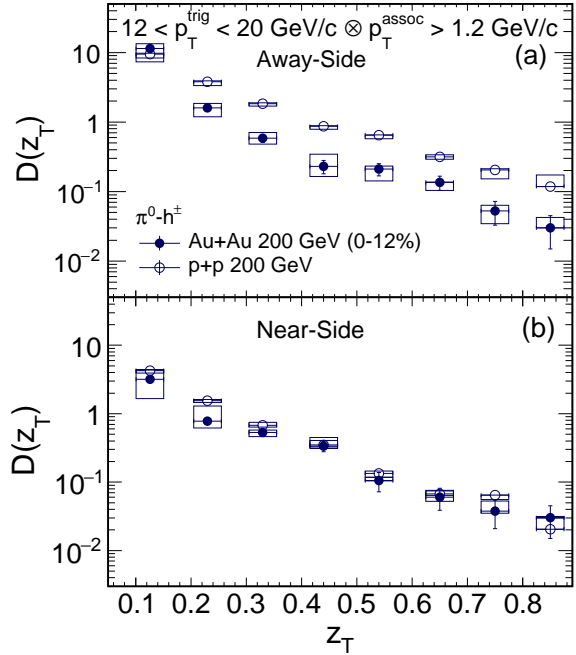}
  \caption{(Color online.) The \zT~dependence of \piZro-$h^{\pm}$
    away-side (a) and near-side (b) associated charged-hadron yields
    per trigger for Au+Au at 0-12$\%$ centrality
  (filled symbols) and  $p+p$ (open symbols) collisions at \sNN~= 200 GeV. Vertical lines represent the statistical
  errors, and the vertical extent of the boxes represents systematic uncertainties. }
\label{Fig2}
\end{center}
\end{figure}

In this measurement, both \piZro~and \DirPho~triggers are required to be within a range of 12 $<$ \ptTrig $<$ 20 \GeVc, or 8 $<$ \ptTrig $<$ 20  \GeVc~for the study of the \ptTrig~dependence.
In contrast to a \DirPho~trigger, a \piZro~trigger carries a fraction of the initial parton energy of the hard-scattered parton.
In this case, the \zT~for a trigger+associated-particle pair is only a loose approximation of the fractional parton energy carried by the jet constituent.
 The integrated away-side and near-side charged-hadron yields per \piZro~trigger, $D(z_{T})$, are plotted as a function of
\zT, both for Au+Au (0-12$\%$ centrality) and $p+p$
collisions, in Fig.~\ref{Fig2}. 
 Yields of the away-side associated charged hadrons are suppressed, in Au+Au relative to $p+p$, at all \zT~except in the low \zT~region. 
On the other hand, no suppression is observed on the near-side in Au+Au, relative to $p+p$ collisions, due to the surface bias 
imposed by triggering on a high-$p_T$ \piZro.\\

\begin{figure}[htbp]
\begin{center}
	 \includegraphics[width=0.5\textwidth]{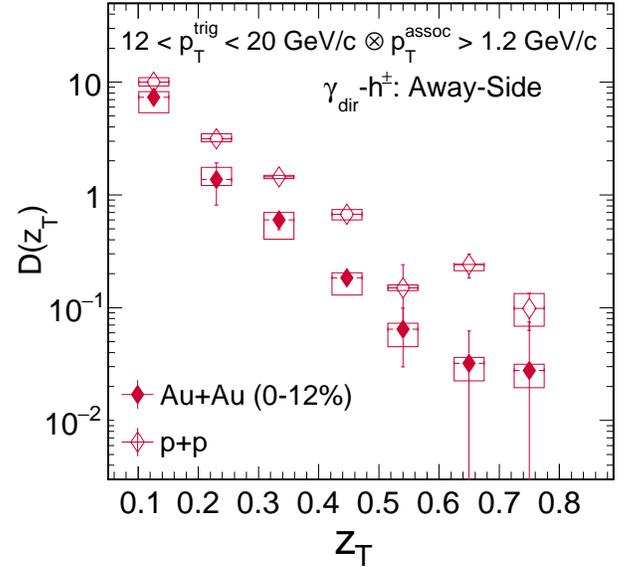}		
  \caption{(Color online.) The \zT~dependence of 
  \DirPho-$h^{\pm}$ away-side associated charged-hadron yields per
  trigger for Au+Au at 0-12$\%$ centrality (filled diamonds) and
  $p+p$ (open diamonds) collisions. Vertical lines represent statistical
  errors, and the vertical extent of the boxes represents systematic uncertainties.}

 \label{Fig3}
\end{center}
\end{figure}

\begin{figure}[htbp]
\begin{center}
 \includegraphics[width=0.5\textwidth]{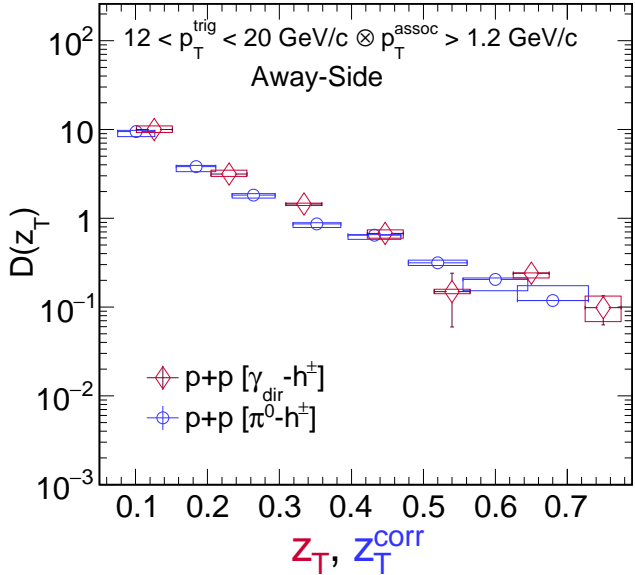}	
  \caption{(Color online.) The \zT~= \ptAssoc/\pT$^{\gamma_{dir}}$ dependences of 
  \DirPho-$h^{\pm}$ away-side associated charged-hadron yields per
  trigger for $p+p$ (open circles) collisions and that of \zT$^{\mathrm{corr}}$ = \ptAssoc/\pT$^{jet}$ dependence of the 
  \piZro-$h^{\pm}$ away-side associated charged-hadron yields (open diamonds) are shown.
  Vertical lines represent statistical
  errors bars, and the vertical extent of the boxes represents systematic uncertainties.}  
 \label{Fig4}
\end{center}
\end{figure}

Figure~\ref{Fig3} shows the away-side $D(z_{T})$
for \DirPho~triggers, as extracted from Eq.~\ref{eq:Yield_dirPhoton}, as a function of \zT~for central Au+Au and minimum-bias $p+p$
collisions.  
The \piZro-triggered away-side charged-hadron yields cannot be directly compared to those of \DirPho~triggers, as the
\piZro~trigger is a fragment of a higher energy parton. 
One can approximate
the fraction of additional energy by integrating \zT~times a
fit to the near-side $D(z_T)$ distribution, measured in $p+p$
collisions, over all \zT~($z_T = 0 \rightarrow \infty$).
The value of that fraction is  
\begin{equation}
\frac{\sum p_T^{assoc}}{p_T^{trig}} = 0.17 \pm 0.04.
\end{equation}
From that, the fraction of energy carried by the $\pi^0$ trigger, with $p_T^{trig}= 12 - 20 $~\GeVc, is estimated to be 
\begin{equation}
\frac{p_T^{trig}}{p_T^{jet-charged}}= 85 \pm 3\%, 
\end{equation}
where $p_T^{jet-charged}$ is equal to the $p_T^{trig}$ plus the total 
$p_T$ carried by the near-side associated charged hadrons.
This is consistent with what is obtained when applying the same analysis on \piZro-triggered 
charged-hadron correlations from a PYTHIA simulation.  In PYTHIA, the neutral associated 
energy can also be accounted for, giving us an estimate of the 
fractional energy carried by the \piZro~trigger, 
when accounting for all associated particles (charged and neutral), 
\begin{equation}
\frac{p_T^{trig}}{p_T^{jet}} = 80 \pm 5\%.
\end{equation}
Applying this ratio as a correction factor to the \zT~values of the 
away-side $D(z_T)$ for \piZro~triggers in $p+p$ collisions 
results in the $D(z_T^{corr})$ distribution, where 
\begin{equation}
z_T^{corr} = \frac{p_T^{assoc}}{p_T^{jet}}.
\end{equation}
Since $z_T^{corr}$ represents the fractional momentum of the jet carried by the associated particles, 
it is (to the extent that the $p_T^{\gamma}$ is a good approximation of the initial \pT~of the recoil parton) 
equivalent to the $z_T$ measured when using \DirPho~triggers.
$D(z_T^{corr})$ is directly compared to the fragmentation function measured 
via direct-photon triggers in Fig.~\ref{Fig4} and shows reasonable agreement.

\begin{figure}[htbp]
\begin{center}
\includegraphics[width=0.5\textwidth]{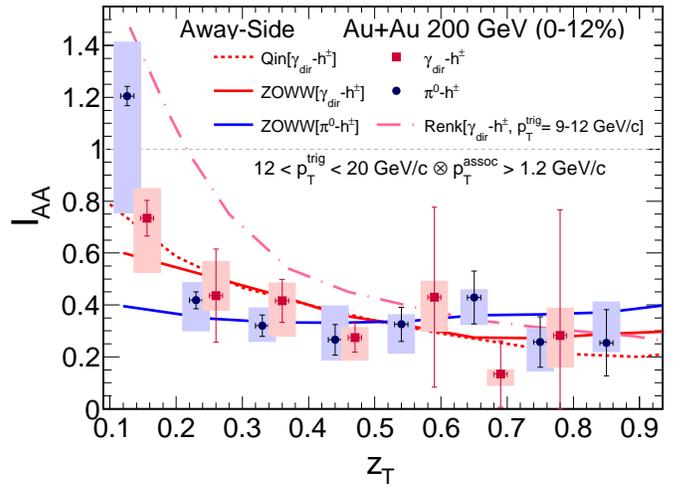}
  
\caption{(Color online.) The  \IAAg~(red
  squares) and \IAApiZro~(blue circles) triggers are plotted as a function of \zT. The points for \IAAg~ are shifted by $+0.03$ in \zT~for visibility.  The vertical lines represent statistical
  error and the vertical extent of the boxes represent systematic errors. The curves
  represent theoretical model predictions~\cite{Qin,Zhang_Owens_EWang_XWang_PRL,Wang,YAJEM}. }
\label{Fig5}
\end{center}
\end{figure}

In order to quantify the medium modification for \DirPho- and \piZro-triggered
recoil jet production as a
function of \zT, the
ratio, defined as 
\begin{equation}
I_{AA} = \frac{D(z_{T})^{AuAu}}{D(z_{T})^{pp}}, 
\label{eq:IAA}
\end{equation}
of the per-trigger conditional yields in Au+Au to those in $p+p$ collisions is calculated. 
In the absence of medium modifications, $I_{AA}$ is expected to be 
equal to unity.
Figure~\ref{Fig5} shows the away-side medium
modification factor for \piZro~triggers
(\IAApiZro) and \DirPho~triggers (\IAAg), as a function of \zT.
\IAApiZro~and \IAAg~show similar suppression
within uncertainties. At low \zT~(0.1 $<$\zT$<$0.2), both \IAApiZro~and
\IAAg~show an indication of less suppression than at higher \zT.  
This observation is not significant in the
\zT-dependence of $I_{AA}$ because the uncertainties in the lowest \zT~bin are large.  However, when $I_{AA}$ is
plotted vs. \ptAssoc~(shown in a later figure), the conclusion is 
supported with somewhat more significance. 
At high \zT, both \IAApiZro~and \IAAg~show a factor $\sim 3-5$ suppression.

Theoretical model predictions,
labeled as Qin~\cite{Qin} and ZOWW~\cite{Zhang_Owens_EWang_XWang_PRL,Wang}, using the same kinematic coverage for
\DirPho~triggered away-side charged-hadron yields, are compared to the data. 
In the model by Qin {\it et al.}, the energy loss mechanism 
is incorporated into a thermalized medium for Au+Au collisions with impact
parameters of $0-2.4$~fm by using a full (3+1)-hydrodynamic evolution model
description. Although this model also includes jet-medium photons (photons coming from the interaction of hard partons with the medium~\cite{TRenk_jetQGPPhoton,Sde_jetQGPPhoton}) and fragmentation photons (photons radiating from hard partons~\cite{Sde_jetQGPPhoton}), both of these contribute to \IAAg~mainly at high \zT~and thus do not 
affect our comparison at low to mid \zT.
The calculation by ZOWW also incorporates 
the parameterized parton energy loss into a bulk-medium evolution~\cite{Wang}.
It does not include fragmentation or jet-medium photons, and also describes the
experimental measurement of \IAAg~as a function of \zT~for the top central Au+Au collisions.  
The calculated \IAApiZro~(also by ZOWW) shows a somewhat larger suppression 
than the \IAAg~at low \zT.  
The difference at 
low \zT~between the \IAAg~and the \IAApiZro~(as calculated by ZOWW) 
is likely due to the color factor effect and the differences in average path 
lengths between \piZro~triggers and \DirPho~triggers.  
The calculated difference in the suppression is approximately 50\% at \zT=0.1. 
The data are not sensitive to this difference within the measured 
uncertainties.
These models (Qin and ZOWW) do not include 
a redistribution of the lost energy to the lower $p_T$ jet fragments, in
contrast to the YaJEM model~\cite{YAJEM}.
The YaJEM model is also shown in Fig.~\ref{Fig5}, although for a somewhat lower trigger \pT~range of 9--12~\GeVc.  
It predicts \IAAg~= 1 at \zT$=0.2$ (corresponding to \ptAssoc $\sim 1.8$~\GeVc)
and rising well above 1 in the \zT~range of 0.1--0.2~\cite{YAJEM}.  
This is calculated with a small integration window of 
$\pi/5$ around $\Delta\phi=\pi$.  
Although this calculation has a different \ptTrig~cut, 
such a large rise is not observed in our data.
In contrast to the other calculations shown (Qin and ZOWW), 
the rise in \IAAg~at low \zT~in YaJEM is predominantly due 
to the redistribution of lost energy.
In this picture, the in-medium shower is modified by the medium and a suppression at high \zT~results 
in an enhancement at lower \zT.  
The authors compare the ``medium-modified shower'' picture 
to an ``energy loss'' picture,
where the energy is carried through the medium by a single parton, and the 
lost energy would only show up at 
extremely low energies and large angles.  
In such a picture, they argue that the 
rise in \IAAg~at low \zT~would be more modest and \IAAg~would remain less than 1.

\begin{figure}[htbp]
\begin{center}
 \includegraphics[width=0.5\textwidth]{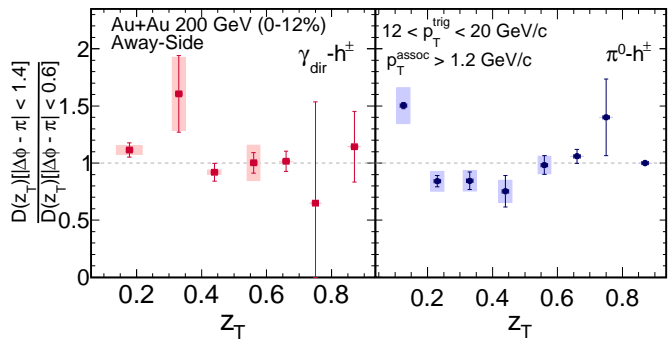}	
  
\caption{(Color online.) The ratios of $D(z_{T})$ obtained using an integration window of $|\Delta\phi - \pi| < 1.4$~radians over that of $|\Delta\phi - \pi| < 0.6$~radians for $\gamma_{dir}-h^{\pm}$ (left panel) and  $\pi^{0}-h^{\pm}$ (right panel), are plotted
  as a function of \zT~for Au+Au at $0-12\%$ central collisions.  The vertical lines represent statistical
  error bars and boxes represent systematic errors.  Since this is a ratio of yields for two overlapping angular windows, much of the uncertainties cancel; and only the surviving uncertainties are shown.}
\label{Fig6}
\end{center}
\end{figure}

Because PHENIX has reported an enhancement at low \zT~(\zT$<0.4$) in \IAAg~at large angles~\cite{PHENIX_GJet}, it is interesting to compare our results over the full integration window of $|\Delta\phi - \pi| < 1.4$~radians to an \IAAg~calculated with a smaller window of 
$|\Delta\phi - \pi| < 0.6$~radians in Fig.~\ref{Fig6}.  Within our uncertainties,
an enhancement effect is only seen in the lowest $z_T$ bin for \piZro~triggers.
However, for the PHENIX measurement, $z_T<0.4$ corresponds to lower \pT~for the associated hadrons
(\raisebox{-0.6ex}{$\stackrel{<}{\sim}$} 2~\GeVc), since the $p_T^{\gamma}$ was chosen in the range of 5--9~\GeVc.  In our analysis, 
associated hadrons with \pT~$<2$~\GeVc~are only present at \zT~$<0.2$.  
The apparent inconsistency between STAR and PHENIX, when investigating 
the recovery of the lost energy as a function of \zT, indicates that \ptAssoc~may be the more pertinent variable. 
The conclusion is that
the ``modified fragmentation function'' (constructed 
from the in-medium jet-like yields as a function of $z_T$) is not universal.  
In particular, the lost energy is not recovered at a fixed range of $z_T$, but perhaps at a given range of $p_T^{assoc}$. 
The conclusion that the lost
energy is recovered at larger angles only for \pT~$<2$~\GeVc, 
regardless of the trigger energy, is consistent with the conclusion of the
STAR paper on jet-hadron correlations~\cite{STAR_jethadron}.

\begin{figure}[htbp]
\begin{center}
\includegraphics[width=0.5\textwidth]{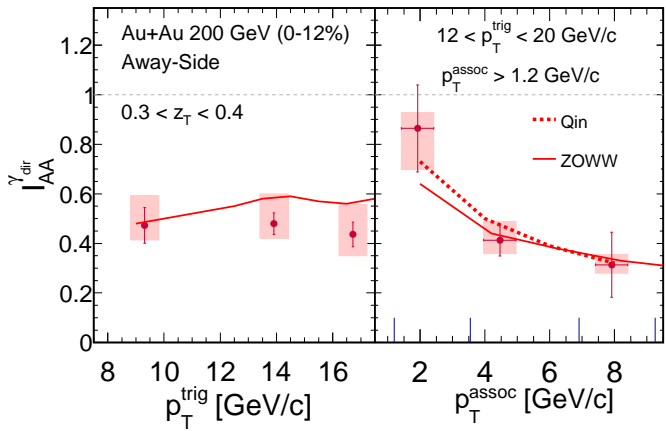}
\caption{(Color online.) The values of \IAAg~are plotted as a function of \ptTrig~(left panel) and \ptAssoc~(right panel). The vertical line and shaded boxes represents statistical and systematic errors, respectively. The curves represent model predictions~\cite{Qin,Zhang_Owens_EWang_XWang_PRL,Wang}.}
\label{Fig7}
\end{center}
\end{figure}

The earlier measurements~\cite{STAR_GJet} at low trigger energy ($8 <$ \ptTrig $< 16$~\GeVc) show the same level of suppression
(factor $3-5$) via the medium
modification factor (\IAApiZro~and \IAAg) down to \zT~$\sim 0.3$.
This suggests that $I_{AA}$ does not depend on the trigger energy at mid to 
high \zT~for
\DirPho~and \piZro-triggered away-side jets with trigger \pT~ranging from 8 to 20~\GeVc. 
This is further investigated in Fig.~\ref{Fig7} with \DirPho~triggers, since 
the photon trigger energy closely approximates the 
initial outgoing parton energy.  The left panel
shows \IAAg~as a function of \ptTrig,
for $0.3 <$ \zT~$< 0.4$.  The per-trigger nuclear modification factor of
\DirPho-triggered away-side charged-hadron yields is independent of the trigger energy of the
\DirPho~within our 25$\%$ systematic uncertainty. This indicates that the away-side parton energy loss is not sensitive to the initial parton energy in this range of 8-20~\GeVc, as measured with our level of precision.
The ZOWW calculation also predicts \IAAg~as a function of \ptTrig~to be approximately flat in this range.
In the right panel, the values of \IAAg~are plotted as function of \ptAssoc. It shows that the low-\ptAssoc~hadrons on
the away-side are not as suppressed as those at high \ptAssoc.  Both model predictions shown~\cite{Qin,Zhang_Owens_EWang_XWang_PRL}, which do not include the redistribution of lost energy, are in agreement with the data.


\section{Summary}
\label{summary}
In summary, in order to understand the medium modification of partons in the QGP, away-side
charged-hadron yields for \DirPho~and \piZro~triggers in
central (0-12$\%$) Au+Au collisions are compared with those in minimum-bias 
$p+p$ collisions.
Both \IAApiZro~and \IAAg~show
similar levels of suppression, with the expected differences due to the 
color-factor effect and the path-length 
dependence not manifesting themselves within experimental uncertainties. 
At low \zT~and low \ptAssoc, the data are 
consistent with less suppression than at higher \ptAssoc.
The suppression shows little difference for integration windows of $\pm$~0.6 vs. $\pm$~1.4~radians around $\Delta\phi = \pi$, with an enhancement at large angles observed only for \zT~$<0.2$ (\ptAssoc~$< 2.4$~ \GeVc) for \piZro~triggers.  
There is no trigger-energy dependence observed in the suppression of \DirPho-triggered yields, 
suggesting little dependence for energy loss on the initial parton energy, in the range of \ptTrig~$=8-20$~\GeVc.
The data are consistent with model calculations ~\cite{Qin,Zhang_Owens_EWang_XWang_PRL,Wang}, in
which the suppression is caused by parton energy loss in a thermalized medium. These calculations do not include redistribution of energy within the shower.  The very large \IAAg~at low \zT~predicted by models of in-medium shower modification (including energy redistribution)~\cite{YAJEM} is not observed for \ptTrig~$>12$~\GeVc.  This is in 
contrast to the PHENIX result~\cite{PHENIX_GJet}, where the \IAAg~exceeds 
unity, for \ptTrig~$5-9$~\GeVc.  However, it is not clear that the redistribution of lost energy would scale with the jet energy. In fact, our studies support previous conclusions that the lost energy reappears predominantly at low \pT~(approximately \pT~$< 2$~\GeVc), regardless of the trigger \pT. This leads to the important conclusion that the modified fragmentation function is not universal ({\it i.e.} it does not have the same $z_T$ dependence for all trigger \pT).
  
\medskip
We thank X. N. Wang and G.-Y Qin for providing their model predictions and helpful discussion.
We thank the RHIC Operations Group and RCF at BNL, the NERSC Center at LBNL, the KISTI Center in
Korea, and the Open Science Grid consortium for providing resources and support. This work was 
supported in part by the Office of Nuclear Physics within the U.S. DOE Office of Science,
the U.S. NSF, the Ministry of Education and Science of the Russian Federation, NSFC, CAS,
MoST and MoE of China, the National Research Foundation of Korea, NCKU (Taiwan), 
GA and MSMT of the Czech Republic, FIAS of Germany, DAE, DST, and UGC of India, the National
Science Centre of Poland, National Research Foundation, the Ministry of Science, Education and 
Sports of the Republic of Croatia, and RosAtom of Russia.

\end{document}